\documentclass[prb,superscriptaddress,aps,epsfig,twocolumn]{revtex4} 

\usepackage{amsfonts}
\usepackage[dvips]{graphicx}
\usepackage{mathrsfs}
\usepackage{amsthm,amssymb}
\usepackage[intlimits]{amsmath}
\usepackage[colorlinks, citecolor=blue, linkcolor=blue, linktocpage]{hyperref}
\usepackage{graphicx}  % needed for figures
\usepackage{dcolumn}   % needed for some tables
\usepackage{bm}        % for math
\usepackage{amssymb}   % for math
\usepackage{floatrow}
\usepackage{mathtools}
\usepackage{braket}
\usepackage[caption=false]{subfig} % for subfigures

\usepackage[english]{babel}
\usepackage{blindtext}
\usepackage{makecell}
\theoremstyle{definition}

\makeatletter
\def\blfootnote{\xdef\@thefnmark{}\@footnotetext}
\makeatother

\newcommand{\one}{\mathbf{1}}

\newcommand{\mP}{\mathcal{P}}

\newcommand{\mN}{\mathcal{N}}
\newcommand{\mM}{\mathcal{M}}
\newcommand{\Z}{\mathbb{Z}}

\begin{document}

\title{Quantum Convolutional Neural Networks}
\author{Iris Cong}
\affiliation{Department of Physics, Harvard University, Cambridge, Massachusetts 02138, USA}
\author{Soonwon Choi}
\email{soonwon@berkeley.edu}
\affiliation{Department of Physics, Harvard University, Cambridge, Massachusetts 02138, USA}
\affiliation{Department of Physics, University of California, Berkeley, CA 94720, USA}
\author{Mikhail D. Lukin}
\affiliation{Department of Physics, Harvard University, Cambridge, Massachusetts 02138, USA}

\begin{abstract}
We introduce and analyze a novel quantum machine learning model motivated 
by convolutional neural networks. Our quantum convolutional neural network (QCNN) makes use of only $O(\log(N))$ variational parameters for input sizes of $N$ qubits, allowing for its efficient training and implementation on realistic, near-term quantum devices. The QCNN architecture combines the multi-scale entanglement renormalization ansatz and quantum error correction.  We explicitly illustrate its potential with two examples. First, QCNN is used  to accurately recognize quantum states associated with  1D symmetry-protected topological phases. We numerically demonstrate that a QCNN trained on a small set of exactly solvable points can reproduce the phase diagram over the entire parameter regime and also provide an exact, analytical QCNN solution. As a second application, we utilize QCNNs to devise a quantum error correction scheme optimized for a given error model. We provide a generic framework to simultaneously optimize both encoding and decoding procedures and find that the resultant scheme significantly outperforms known quantum codes of comparable complexity. Finally, potential experimental realization and generalizations of QCNNs are discussed. 
\end{abstract}
\maketitle

Machine learning  based on  neural networks has recently provided significant advances for many practical applications\cite{LeCun15}.  In physics, one  natural application involves the study of quantum many-body systems, where the extreme complexity of many-body states often makes theoretical analysis intractable. This has led to a number of recent works using machine learning to study properties of quantum systems\cite{Carleo17,Nieuwenburg17,Carrasquilla17,Wang16,Levine19,YZhang17}, using physical concepts to interpret machine learning\cite{Lin17,Mehta14}, or using quantum computers to enhance conventional machine learning tasks\cite{Biamonte17,Dunjko16,Farhi18,Huggins18}. 

In this work, motivated by the progress towards realizing quantum information processors\cite{Ladd10,Monroe13,Devoret13,Awschalom13}, we bridge these approaches by proposing a quantum circuit model inspired by machine learning and illustrating its success for two important classes of quantum many-body problems.
The first class of problems we consider is
{\it quantum phase recognition} (QPR), which asks whether a given input quantum state $\rho_\textrm{in}$ belongs to a particular quantum phase of matter. Critically, in contrast to many existing schemes based on tensor network descriptions \cite{Huang13,Singh13,Kim17}, we assume $\rho_\textrm{in}$ is prepared in a physical system without direct access to its classical description. 
The second class, {\it quantum error correction (QEC) optimization}, asks for an optimal QEC code for a given, {\it a priori} unknown error model such as dephasing or potentially correlated depolarization in realistic experimental settings.

The highly complex and intrinsically quantum nature of these problems makes them particularly difficult to solve using existing classical and quantum machine learning techniques. While conventional machine learning with large-scale neural networks can successfully solve analogous classical problems such as image recognition or improving classical error correction\cite{LeCun15}, the exponentially large many-body Hilbert space hinders efficiently translating such quantum problems into a classical framework without performing exponentially difficult quantum state or process tomography\cite{Haah17}. Quantum algorithms avoid this overhead, but the limited size and coherence times of near-term quantum devices prevent the use of large-scale networks; thus, it is vital to first theoretically understand the most important machine learning mechanisms that must be implemented. In this work, we introduce a quantum machine learning method for QPR and QEC optimization, provide both theoretical insight and numerical demonstrations for its success, and show its feasibility for near-term experimental implementation.

\begin{figure}
\includegraphics[width=0.96\textwidth]{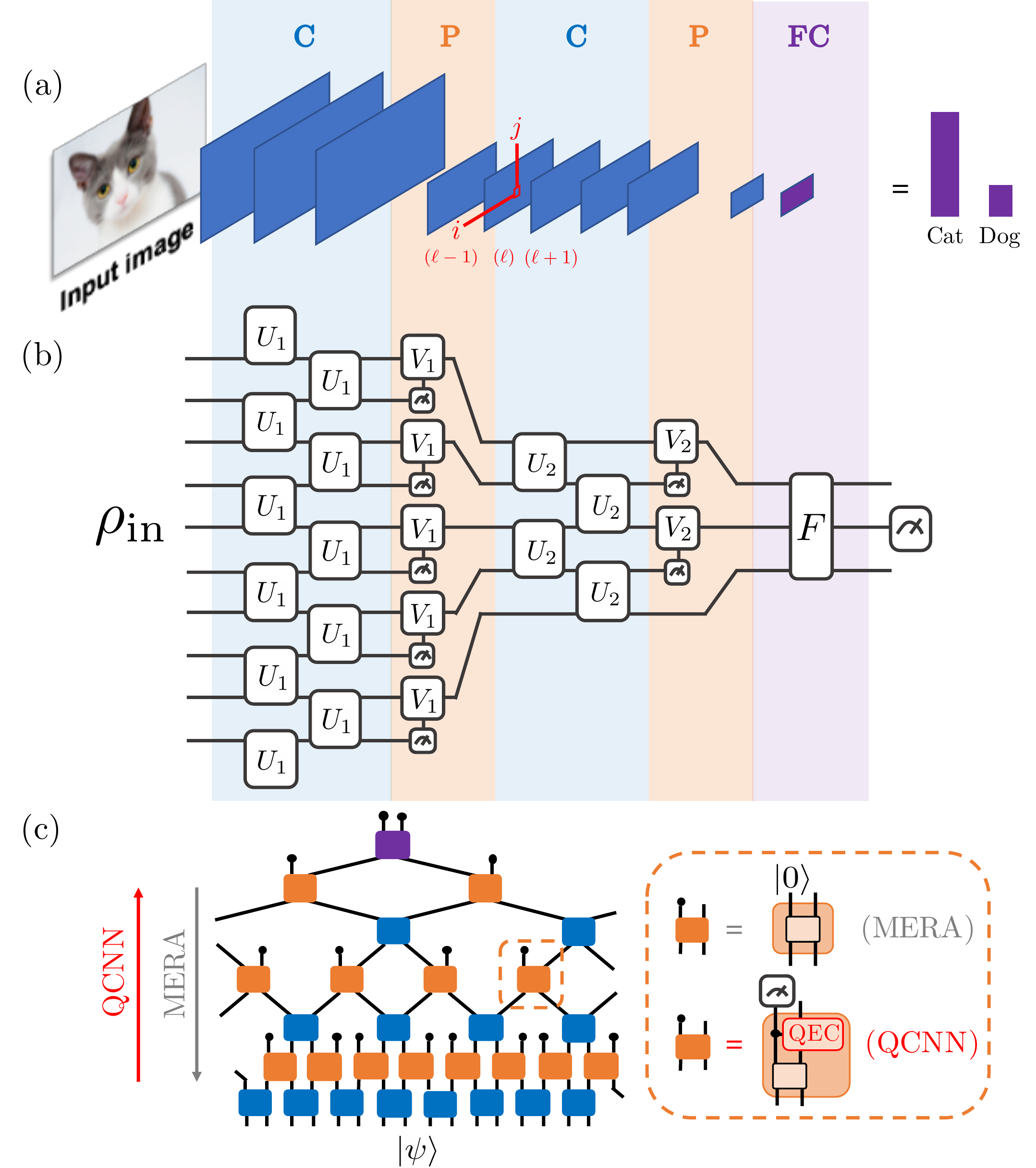}
    \caption{(a) Simplified illustration of CNNs. A sequence of image processing layers---convolution (C), pooling (P), and fully connected (FC)---transforms an input image into a series of feature maps (blue rectangles), and finally into an output probability distribution (purple bars).  (b) QCNNs inherit a similar layered structure. (c) QCNN and MERA share the same circuit structure, but run in reverse directions.}
\end{figure}

\section*{QCNN CIRCUIT MODEL}

Convolutional neural networks (CNNs) provide a successful machine learning architecture for classification tasks such as image recognition~\cite{LeCun95,Krizhevsky12,LeCun15}. A CNN generally consists of a sequence of different (interleaved) layers of image processing; in each layer, an intermediate 2D array of pixels, called a feature map, is produced from the previous one (Figure 1a)\cite{foot1}. The convolution layers compute new pixel values $x_{ij}^{(\ell)}$ from a linear combination of nearby ones in the preceding map $x_{i,j}^{(\ell)} = \sum_{a,b=1}^w w_{a,b} x_{i+a,j+b}^{(\ell -1)}$, where the weights $w_{a,b}$ form a $w \times w$ matrix.
Pooling layers reduce feature map size, e.g. by taking the maximum value from a few contiguous pixels, and are often followed by application of a nonlinear (activation) function. Once the feature map size becomes sufficiently small, the final output is computed from a function that depends on all remaining pixels (fully connected layer). The weights and fully connected function are optimized by training on large datasets. In contrast, variables such as the number of convolution and pooling layers and the size $w$ of the weight matrices (known as hyperparameters) are fixed for a specific CNN \cite{LeCun15}. CNN's key properties are thus translationally invariant convolution and pooling layers, each characterized by a constant number of parameters (independent of system size), and sequential data size reduction (i.e., a hierarchical structure). 

Motivated by this architecture we  introduce a quantum circuit model extending these key properties to the quantum domain (Fig.~1b). The circuit's input is an unknown quantum state $\rho_{\text{in}}$. A convolution layer applies a single quasi-local unitary ($U_i$) in a translationally-invariant manner for finite depth. For pooling, a fraction of qubits are measured, and their outcomes determine unitary rotations ($V_j$) applied to nearby qubits. Hence, nonlinearities in QCNN arise from reducing the number of degrees of freedom. Convolution and pooling layers are performed until the system size is sufficiently small; then, a fully connected layer is applied as a unitary $F$ on the remaining qubits. Finally, the outcome of the circuit is obtained by measuring a fixed number of output qubits. As in the classical case, circuit structures (i.e. QCNN hyperparameters) such as the number of convolution and pooling layers are fixed, and the unitaries themselves are learned.

A QCNN to classify $N$-qubit input states is thus characterized by $O(\log(N))$ parameters. This corresponds to doubly exponential reduction compared to a generic quantum circuit-based classifier\cite{Farhi18} and allows for efficient learning and implementation. For example, given 
classified training data $\{(\ket{\psi_\alpha},  y_\alpha): \alpha = 1, ..., M\}$, where $\ket{\psi_\alpha}$ are input states and $y_\alpha = 0$ or $1$ are corresponding binary classification outputs, one could compute the mean-squared error
\begin{equation}
\label{eq:mse}
\text{MSE} 
= \frac{1}{2M} \sum_{\alpha = 1}^M (y_i - f_{\{U_i,V_j,F\}}(\ket{\psi_\alpha}))^2.
\end{equation}
Here, $f_{\{U_i,V_j,F\}}(\ket{\psi_\alpha})$ denotes the expected QCNN output value for input $\ket{\psi_\alpha}$.
Learning then consists of initializing all unitaries and successively optimizing them until convergence, e.g. via gradient descent.

To gain physical insight into the mechanism underlying QCNNs and motivate their application to the problems under consideration, we now relate our circuit model to two well-known concepts in quantum information theory---the multiscale entanglement renormalization ansatz\cite{Vidal07} (MERA) and quantum error correction (QEC). The MERA framework provides an efficient tensor network representation of many classes of interesting many-body wavefunctions, including those associated with critical systems~\cite{Vidal07,Aguado08,Pfeifer09}. 
 A MERA can be understood as a quantum state generated by a sequence of unitary and isometry layers applied to an input state (e.g. $\ket{00}$). While each isometry layer introduces a set of new qubits in a predetermined state (e.g. $\ket{0}$) before applying unitary gates on nearby ones, unitary layers simply apply quasi-local unitary gates to the existing qubits (Figure 1c).
This exponentially growing, hierarchical structure allows for the long-range correlations associated with  critical systems. The QCNN circuit has similar structure, but runs in the reverse direction.
Hence, for any given state $\ket{\psi}$ with a MERA representation, there is always a QCNN that recognizes $\ket{\psi}$ with deterministic  measurement outcomes; one such QCNN is simply the inverse of the MERA circuit. 

For input states other than $\ket{\psi}$, however, such a QCNN  does not generally produce deterministic measurement outcomes.
These additional degrees of freedom distinguish QCNN from MERA.
Specifically, we can identify the measurements as syndrome measurements in QEC~\cite{Preskill98}, which determine error correction unitaries $V_j$ to apply to the remaining qubit(s). Thus, a QCNN circuit with multiple pooling layers can be viewed as a combination of MERA --- an important variational ansatz for many-body wavefunctions --- and nested QEC --- a mechanism to detect and correct local quantum errors without collapsing the wavefunction.
This makes QCNN a powerful architecture to classify input quantum states or devise novel QEC codes.
In particular, for QPR, the QCNN can provide a MERA realization of a representative state $\ket{\psi_0}$ in the target phase. Other input states within the same phase can be viewed as $\ket{\psi_0}$ with local errors, which are repeatedly corrected by the QCNN in multiple layers.
In this sense, the QCNN circuit can mimic renormalization-group (RG) flow, a methodology which successfully classifies many families of quantum phases~\cite{Sachdev11}. For QEC optimization, the QCNN structure allows for simultaneous optimization of efficient encoding and decoding schemes with potentially rich entanglement structure.

\begin{figure*}
\includegraphics[width=0.6\textwidth]{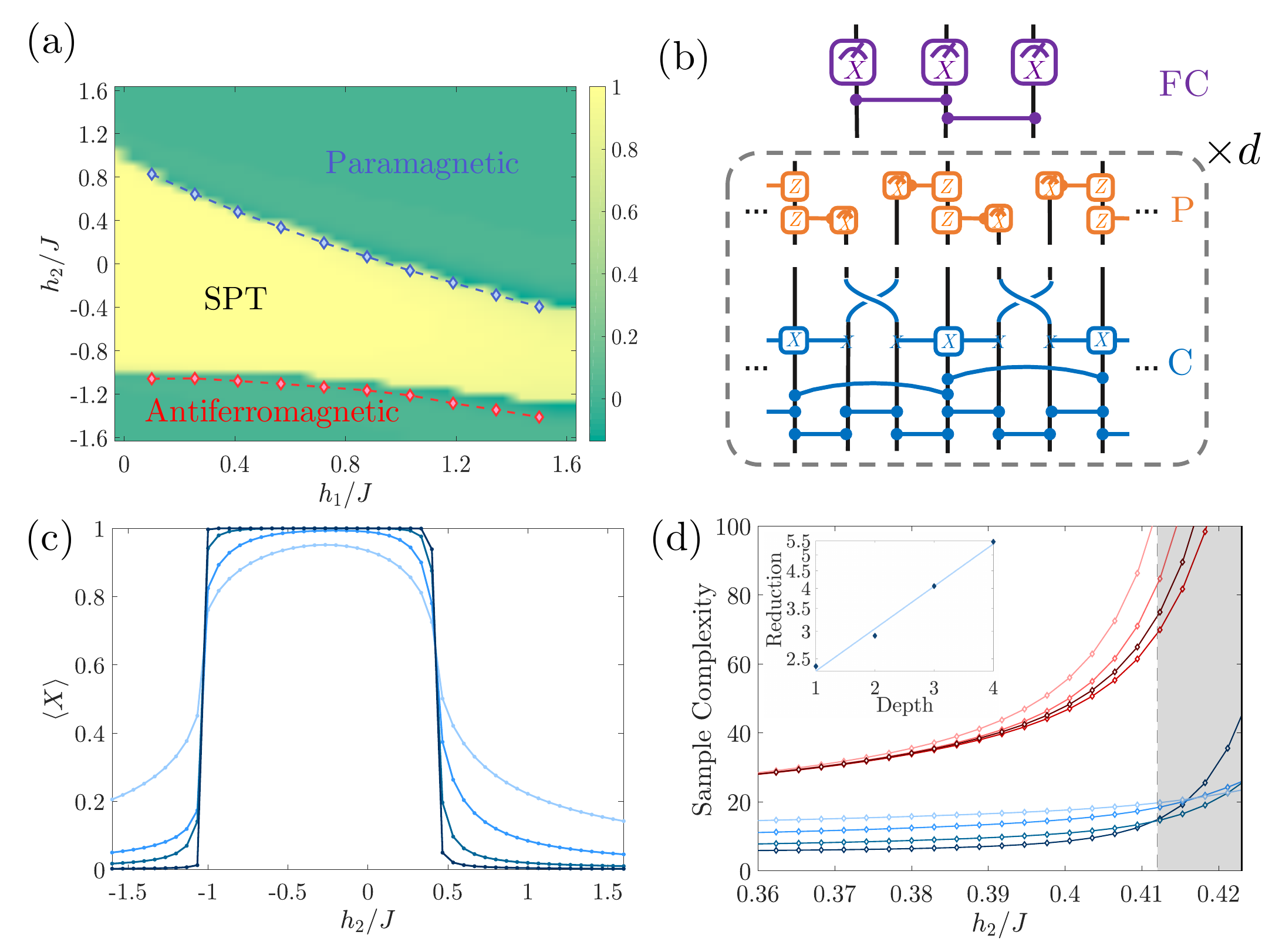}
\caption{(a) The phase diagram of the Hamiltonian in the main text. The phase boundary points (blue and red diamonds) are extracted from infinite size DMRG numerical simulations, while the color represents the output from the exact QCNN circuit for input size $N = 45$ spins (see Methods).
(b) Exact QCNN circuit to recognize a $\Z_2 \times \Z_2$ SPT phase.
Blue line segments represent controlled-phase gates, blue three-qubit gates are Toffoli gates with the control qubits in the $X$ basis, and orange two-qubit gates flip the target qubit's phase when the $X$ measurement yields $-1$.
The fully connected layer applies controlled-phase gates followed by an $X_i$ projection, effectively measuring $Z_{i-1}X_i Z_{i+1}$.
(c) Exact QCNN output along $h_1 = 0.5J$ for $N = 135$ spins, $d = 1,...,4$. 
(d) Sample complexity of QCNN at depths $d = 1, ... 4$ (blue) versus SOPs of length $N/2$, $N/3$, $N/5$, and $N/6$ (red) to detect the SPT/paramagnet phase transition along $h_1 = 0.5J$ for $N = 135$ spins. The critical point is identified as $h_2/J = 0.423$ using infinite size DMRG (bold line). Darkening colors show higher QCNN depth or shorter string lengths. 
In the shaded area, the correlation length exceeds the system size and finite-size effects can considerably affect our results. Inset: The ratio of SOP sample complexity to QCNN sample complexity is plotted as a function of depth $d$ on a logarithmic scale for $h_1/J = 0.3918$. In the numerically accessible regime, this reduction of sample complexity scales exponentially as $1.73 e^{0.28 d}$ (trendline).}
\label{fig:cluster-qcnn}
\end{figure*}

\section*{DETECTING A 1D SPT PHASE}

We first demonstrate the potential of QCNN explicitly by applying it to QPR in a class of one-dimensional many-body systems.  Specifically, we consider a $\Z_2 \times \Z_2$ symmetry-protected topological (SPT) phase $\mathcal{P}$, a phase containing the $S=1$ Haldane chain\cite{Haldane83}, and ground states $\{\ket{\psi_G}\}$ of a family of Hamiltonians on a spin-$1/2$ chain with open boundary conditions:
\begin{equation}
\label{eq:zxz-ham}
H = -J \sum_{i=1}^{N-2} Z_i X_{i+1} Z_{i+2}-h_1 \sum_{i=1}^{N} X_i - h_2 \sum_{i=1}^{N-1} X_i X_{i+1}.
\end{equation}
$X_i, Z_i$ are Pauli operators for the spin at site $i$, and the $\Z_2 \times \Z_2$ symmetry is generated by $X_\textrm{even(odd)} = \prod_{i \in \textrm{even(odd)}} X_i$. 
Figure \ref{fig:cluster-qcnn}a shows the phase diagram as a function of $(h_1/J,h_2/J)$.
When $h_2 = 0$, the Hamiltonian is exactly solvable via Jordan-Wigner transformation\cite{Sachdev11}, confirming that $\mathcal{P}$ is characterized by nonlocal order parameters. When $h_1 = h_2 = 0$, all terms are mutually commuting, and a ground state is the 1D cluster state. Our goal is to identify whether an given, unknown ground state drawn from the phase diagram belongs to $\mathcal{P}$.
%\subsection*{Exact QCNN Circuit}

As an example, we first present an exact, analytical QCNN circuit that recognizes $\mathcal{P}$, see Figure \ref{fig:cluster-qcnn}b. 
The convolution layers involve controlled-phase gates as well as Toffoli gates with controls in the $X$-basis, and pooling layers perform phase-flips on remaining qubits when one adjacent measurement yields $X = -1$. This convolution-pooling unit is repeated $d$ times, where $d$ is the QCNN depth. The fully connected layer measures $Z_{i-1}X_iZ_{i+1}$ on the remaining qubits.
Figure \ref{fig:cluster-qcnn}c shows the QCNN output for a system of $N = 135$ spins and $d = 1,...,4$ along $h_2 = 0.5 J$, obtained using matrix product state simulations. As $d$ is increased, the measurement outcomes show sharper changes around the critical point, and the output of a $d=2$ circuit already reproduces the phase diagram with high accuracy (Figure \ref{fig:cluster-qcnn}a).
This QCNN can also be used for other Hamiltonian models belonging to the same phase, such as the $S = 1$ Haldane chain\cite{Haldane83} (see Methods).

\subsection*{Sample Complexity}
The performance of a QPR solver can be quantified by sample complexity\cite{Haah17}: what is the expected number of copies of the input state required to identify its quantum phase?
We demonstrate that the sample complexity of our exact QCNN circuit is significantly better than that of conventional methods.
In principle, $\mathcal{P}$ can be detected by 
measuring a nonzero expectation value of string order parameters (SOP)\cite{Haegeman12,Pollman12} such as
\begin{equation}
\label{eq:cluster-string-op}
\mathcal{S}_{ab} = Z_a X_{a+1} X_{a+3} ... X_{b-3} X_{b-1} Z_b.
\end{equation}
In practice, however, the expectation values of SOP vanish near the phase boundary due to diverging correlation length\cite{Pollman12}; since quantum projection noise is maximal in this vicinity, many experimental repetitions are required to affirm a nonzero expectation value. 
In contrast, the QCNN output is much sharper near the phase transition, so fewer  repetitions are required. 

Quantitatively, given some $\ket{\psi_{\text{in}}}$ and SOP $S$, a projective measurement of $S$ can be modeled as a (generalized) Bernoulli random variable, where the outcome is 1 with probability $p = (\bra{\psi_\text{in}} S \ket{\psi_{\text{in}}}+1)/2$ and $-1$ with probability $1-p$ (since $S^2 = \mathbf{1}$); after $M$ binary measurements, we estimate $p$. $p > p_0 = 0.5$ signifies $\ket{\psi_{\text{in}}} \in \mP$. We define the sample complexity $M_{\text{min}}$ as the minimum $M$ to test whether $p > p_0$ with 
95\% confidence using an arcsine variance-stabilizing transformation\cite{Brown01}:
\begin{equation}
\label{eq:sample-complexity}
M_{\text{min}} = \frac{1.96^2}{(\arcsin{\sqrt{p}-\sqrt{\arcsin{p_0}}})^2}.
\end{equation}
\noindent
Similarly, the sample complexity for a QCNN can be determined by replacing $\bra{\psi_\text{in}} S \ket{\psi_{\text{in}}}$ by the QCNN output expectation value in the expression for $p$. 

Figure \ref{fig:cluster-qcnn}d shows the sample complexity for the QCNN at various depths and SOPs of different lengths. Clearly, QCNN requires substantially fewer input copies throughout the parameter regime, especially near criticality. More importantly, although the SOP sample complexity scales independently of string length, the QCNN sample complexity consistently improves with increasing depth and is only limited by finite size effects in our simulations. In particular, compared to SOPs, QCNN reduces sample complexity by a factor which scales exponentially with the QCNN's depth in numerically accessible regimes (inset).
Such scaling arises from the iterative QEC performed at each depth and is not expected from any measurements of simple (potentially nonlocal) observables. We show in Methods that our QCNN circuit measures a {\it multiscale} string order parameter---a sum of products of exponentially many different SOPs which remains sharp up to the phase boundary.

\begin{figure}
           \includegraphics[width=0.85\textwidth]{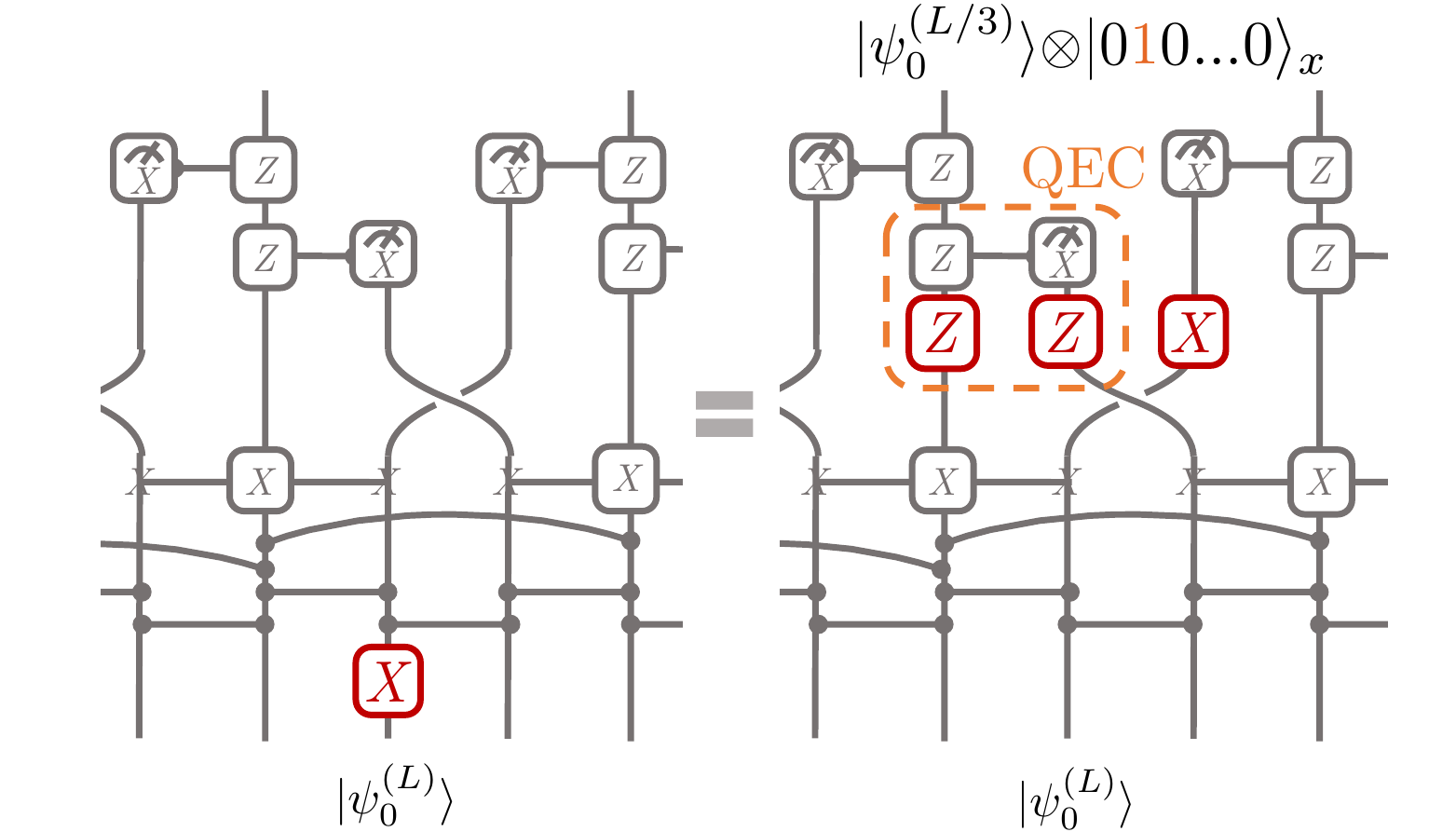}
    \caption{Action of cluster model QCNN convolution-pooling unit on a state with a single-qubit $X$ error.}
    \label{fig:cluster-qcnn-qec}
\end{figure}

\subsection*{MERA and QEC}
Additional insights into the QCNN's performance are revealed by interpreting it in terms of MERA and QEC.
In particular, our QCNN is specifically designed to contain the MERA representation of the 1D cluster state ($\ket{\psi_0}$)---the ground state of $H$ with $h_1 = h_2 = 0$---such that it becomes a stable fixed point. When $\ket{\psi_0}$ is fed as input, each convolution-pooling unit produces the same state $\ket{\psi_0}$ with reduced system size in the unmeasured qubits, while yielding deterministic outcomes ($X=1$) in the measured qubits. 
The fully connected layer measures the SOP for $\ket{\psi_0}$.
When an input wavefunction is perturbed away from $\ket{\psi_0}$, our QCNN corrects such ``errors.'' For example, if a single $X$ error occurs, the first pooling layer identifies its location, and controlled unitary operations correct the error propagated through the circuit (Fig. \ref{fig:cluster-qcnn-qec}).
Similarly, if an initial state has multiple, sufficiently separated errors (possibly in coherent superpositions), the error density after several iterations of convolution and pooling layers will be significantly smaller\cite{Zeng16}.
If the input state converges to the fixed point, our QCNN classifies it into the SPT phase with high fidelity.
Clearly, this mechanism resembles the classification of quantum phases based on renormalization-group (RG) flow.

\begin{figure}[t]
\includegraphics[width=0.85\textwidth]{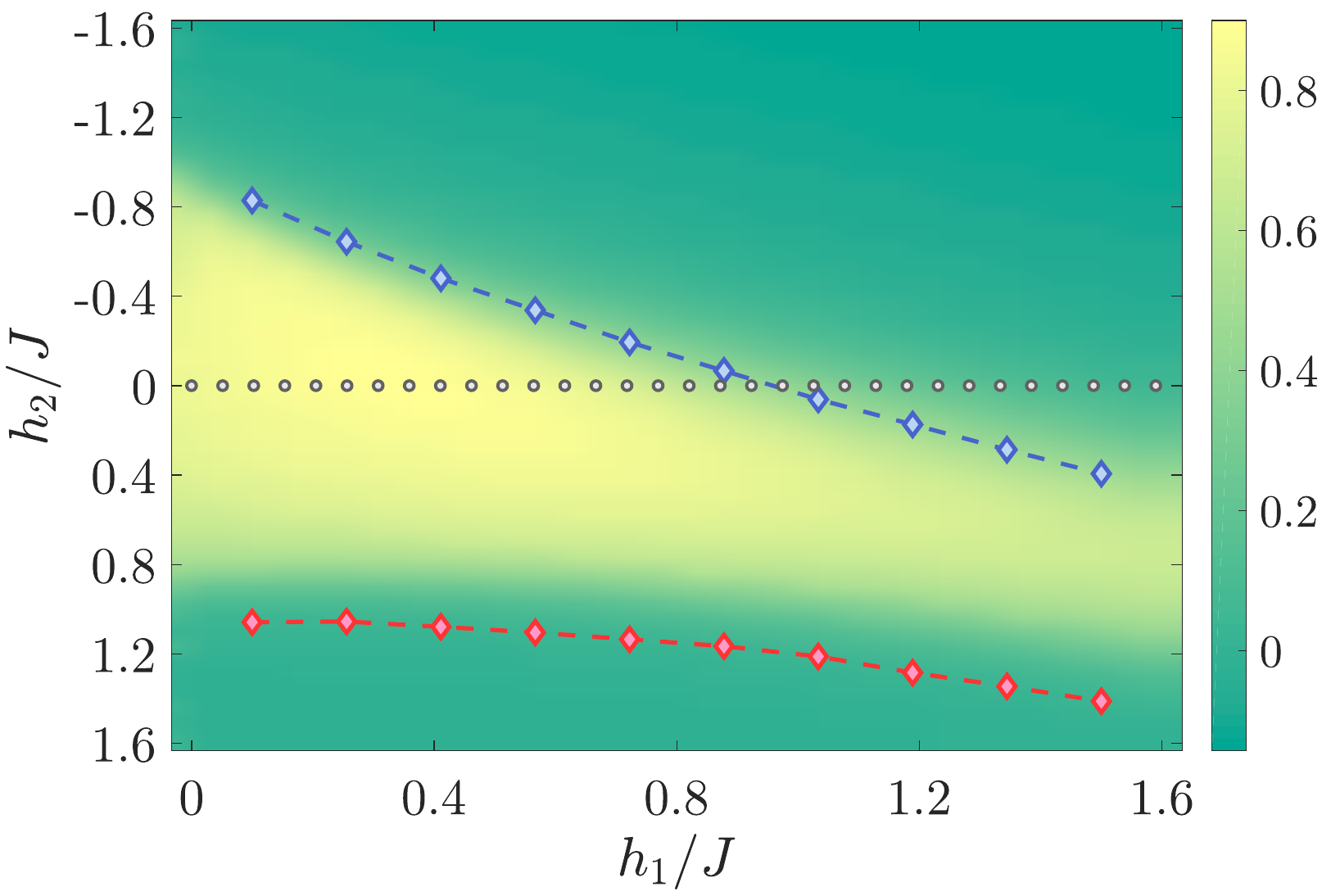}
\label{fig:learning}
\caption{
Output of a randomly initialized and trained QCNN for $N = 15$ spins and depth $d = 1$. Gray dots show (most of)\cite{foot3} the training data points, which are 40 equally spaced points on a line where the Hamiltonian is solvable by Jordan-Wigner transformation ($h_2 = 0, h_1 \in [0,2]$). The blue and red diamonds are phase boundary points extracted from infinite size DMRG numerical simulations, while the colors represent the expectation value of the QCNN output.
}
\end{figure}

\subsection*{Obtaining QCNN from Training Procedure}
Having analytically illustrated the computational power of the QCNN circuit model, we now demonstrate how a QCNN for $\mathcal{P}$ can also be obtained using the learning procedure. 
In our example, the QCNN's hyper-parameters are chosen such that there are four convolution layers and one pooling layer at each depth, followed by a fully connected layer (see Methods). 
Initially, all unitaries are set to random values.
Because classically simulating our training procedure requires expensive computational resources, 
we focus on a relatively small system with $N = 15$ spins and QCNN depth $d = 1$;  there are a total of 1309 parameters to be learned (see Methods).
Our training data consists of 40 evenly spaced points along the line $h_2 = 0$, where the Hamiltonian is exactly solvable by Jordan-Wigner transformation.  Using gradient descent with the mean-squared error function (\ref{eq:mse}), we iteratively update the unitaries until convergence (see Methods). The classification output of the resulting QCNN for generic $h_2$ is shown in Fig. 4.
Remarkably, this QCNN accurately reproduces the 2D phase diagram  over the entire parameter regime, even though the model was trained only on samples from a set of solvable points which does not even cross the lower phase boundary.

This example illustrates how the QCNN structure avoids overfitting to training data with its exponentially reduced number of parameters. While the training dataset for this particular QPR problem consists of solvable points, more generally, 
such a dataset can be obtained by using traditional methods (e.g. measuring SOPs) to classify representative states that can be efficiently generated either numerically or experimentally\cite{Schwarz12,Ge16}.

\begin{figure}[t]
           \includegraphics[width=0.75\textwidth]{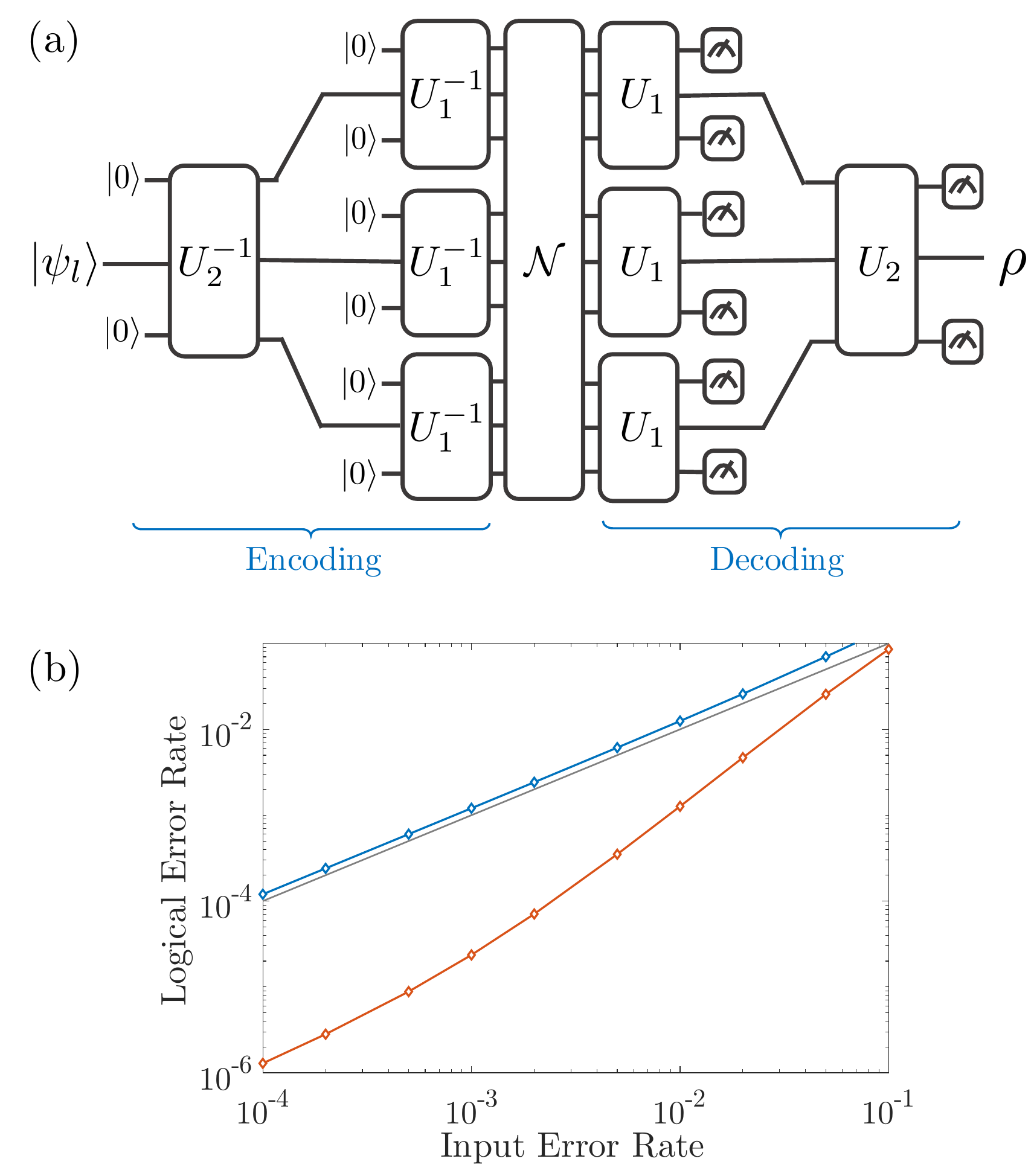}
    \caption{(a) Schematic diagram for using QCNNs to optimize QEC. The inverse QCNN encodes a single logical qubit $\ket{\psi_l}$ into 9 physical qubits, which undergo noise $\mN$. QCNN then decodes these to obtain the logical state $\rho$. Our aim is to maximize $\bra{\psi_l} \rho \ket{\psi_l}$. (b) Logical error rate of Shor code (blue) versus a learned QEC code (orange) in a correlated error model. The input error rate is defined as the sum of all probabilities $p_\mu$ and $p_{xx}$. The Shor code has worse performance than performing no error correction at all (identity, gray line), while the optimized code can still significantly reduce the error rate.}
    \label{fig:qcnn-qec}
\end{figure}

\section*{OPTIMIZING QUANTUM ERROR CORRECTION}
As seen in the previous example, the QCNN's architecture enables one to perform effective QEC.
We next leverage this feature to design a new QEC code itself that is optimized for a given error model.
More specifically, 
any QCNN circuit (and its inverse) can be viewed as a decoding (encoding) quantum channel between the physical input qubits and the logical output qubit. The encoding scheme introduces sets of new qubits in a predetermined state, e.g. $\ket{0}$, while the decoding scheme performs measurements (Fig.~\ref{fig:qcnn-qec}a).
Given a error channel $\mN$, our aim is therefore to maximize the recovery fidelity
\begin{equation}
\label{eq:qcnn-overlap}
f_q = \sum_{\ket{\psi_l} \in \{\ket{\pm x,y,z}\}} \bra{\psi_l}\mM_q^{-1}(\mN(\mM_q(\ket{\psi_l}\bra{\psi_l})))\ket{\psi_l},
\end{equation}
where $\mM_q$($\mM_q^{-1}$) is the encoding (decoding) scheme generated by a QCNN circuit, and $\ket{\pm x,y,z}$ are the $\pm 1$ eigenstates of the Pauli matrices.
Thus, our method simultaneously optimizes both encoding and decoding schemes, while ensuring their efficient implementation in realistic systems. 
Importantly, the variational optimization can be carried out with a unknown $\mN$ since $f_q$ can be evaluated experimentally.

To illustrate the potential of this procedure, we consider a two-layer QCNN with $N = 9$ physical qubits and 126 variational parameters (Figure \ref{fig:qcnn-qec}a and Methods). 
This particular architecture includes the nested (classical) repetition codes\cite{} and the 9-qubit Shor code\cite{Shor95}; in the following, we compare our performance to the better of the two.
We consider three different input error models: (1) independent single-qubit errors on all qubits with equal probabilities $p_\mu$ for $\mu=X$, $Y$, and $Z$ errors or (2) anisotropic probabilities $p_x \neq p_y = p_z$, and (3) independent single-qubit anisotropic errors with additional two-qubit correlated errors $X_i X_{i+1}$ with probability $p_{xx}$.

Upon initializing all QCNN parameters to random values and numerically optimizing them to maximize $f_q$, we find that our model produces the same logical error rate as known codes in case (1), but can reduce the error rate by a constant factor in case (2), depending on the specific input error probability ratios (e.g. 14\% for $p_x = 1.8 p_y$, or 50\% for $p_x = 0.4 p_y$---see Methods). More drastically, in case (3), the optimized QEC code performs significantly better than known codes (Figure \ref{fig:qcnn-qec}b). Specifically, because the Shor code is only guaranteed to correct arbitrary single-qubit errors, it performs even worse than using no error correction, while the optimized QEC code performs much better. This example demonstrates the power of using QCNNs to obtain and optimize new QEC codes for realistic, a priori unknown error models.

\section*{EXPERIMENTAL REALIZATIONS}

Our QCNN architecture can be efficiently implemented on several state-of-the-art experimental platforms.
The key ingredients for realizing QCNNs include the efficient preparation of quantum many-body input states, the application of two-qubit gates at various length scales, and projective measurements\cite{foot4}.
These capabilities have already been demonstrated in multiple programmable quantum simulators consisting of $N\geq 50$ qubits based on trapped neutral atoms and ions, or superconducting qubits~\cite{Bernien17,Zhang17,Brydges18,Harris18}.

As an example, we present a feasible protocol for near-term implementation of our exact cluster model QCNN circuit via neutral Rydberg atoms~\cite{Bernien17,Labuhn16}, where long-range dipolar interactions allow high fidelity entangling gates~\cite{Levine18} among distant qubits in a variable geometric arrangement. The qubits can be encoded in the hyperfine ground states, 
where one of the states can be coupled to the Rydberg level to perform efficient entangling operations via the Rydberg-blockade mechanism~\cite{Levine18}; an explicit implementation scheme for every gate in Fig. 2b is provided in Methods.
 Our QCNN at depth $d$ with $N$ input qubits requires at most $\frac{7N}{2}(1-3^{1-d}) + N 3^{1-d}$ multi-qubit operations and $4d$ single-qubit rotations.
For a realistic effective coupling strength $\Omega \sim 2\pi \times 10-100 \text{ MHz}$ and single-qubit coherence time $\tau \sim 200 \text{ } \mu\text{s}$ limited by the Rydberg state lifetime, approximately $\Omega \tau \sim 2\pi \times 10^3 - 10^4$ multi-qubit operations can be performed, and a $d=4$ QCNN on $N \sim 100$ qubits feasible. 
These estimates are reasonably conservative as we have not considered advanced control techniques such as pulse-shaping\cite{Freeman98}, or potentially parallelizing independent multi-qubit operations.

\section*{OUTLOOK}
These considerations indicate that QCNNs provide a promising quantum machine learning paradigm.  Several interesting generalizations and future directions can be considered. First, while we have only presented the QCNN circuit structure for recognizing 1D phases, it is straightforward to generalize the model to higher dimensions, where phases with intrinsic topological order such as the toric code are supported\cite{Kitaev97,Levin04}. Such studies could potentially identify nonlocal order parameters with low sample complexity for lesser-understood phases such as quantum spin liquids\cite{Savary17} or anyonic chains\cite{Feiguin07}. To recognize more exotic phases, we could also relax the translation-invariance constraints, resulting in $O(N)$ parameters for system size $N$, or use ancilla qubits to implement parallel feature maps following traditional CNN architecture. 
Further  extensions can incorporate optimizations for fault-tolerant operations on QEC code spaces.
Finally, while we have used a finite-difference scheme to compute gradients in our learning demonstrations, the structural similarity of QCNN with its classical counterpart motivates adoption of more efficient schemes such as backpropagation\cite{LeCun15}. 

{\it Acknowledgment.} The authors thank Xiao-Gang Wen, Ignacio Cirac, Xiaoliang Qi, Edward Farhi, John Preskill, Wen Wei Ho, Hannes Pichler, Ashvin Vishwanath, Chetan Nayak, and Zhenghan Wang for insightful discussions. I.C. acknowledges support from the Paul and Daisy Soros Fellowship, the Fannie and John Hertz Foundation, and the Department of Defense through the National Defense Science and Engineering Graduate Fellowship Program. S.C. acknowledges support from the Miller Institute for Basic Research in Science. This work was supported through the National Science Foundation (NSF), the Center for Ultracold Atoms,  the Vannevar Bush Faculty Fellowship, and Google Research Award.

\bibliography{refs}

\begin{thebibliography}{50}
\bibitem[{foo({\natexlab{b}})}]{foot3}
\bibinfo{note}{We included some training data points outside the plotted 2D
  parameter regime (namely those with $h_1 \in [1.6,2]$) to obtain an equal
  number of points inside and outside the SPT phase, and avoid bias in the
  prior distribution.}
\bibitem[{foo({\natexlab{c}})}]{foot4}
\bibinfo{note}{We emphasize that the measurements of intermediate qubits and
  feed-forwarding can be replaced by controlled two-qubit unitary operations so
  that measurements are only performed at the end of an experimental sequence,
  as in stabilizer-based QEC.}
\end{thebibliography}

\newpage

\begin{center}
{\bf \normalsize Methods}
\end{center}

\setcounter{figure}{6}

\section*{Phase Diagram and QCNN Circuit Simulations}

The phase diagram in the main text (Fig. 2a) was numerically obtained using the infinite size density-matrix renormalization group (DMRG) algorithm. 
We generally follow the method outlined in Ref.~\onlinecite{McCulloch08} with the maximum bond dimension 150. To extract each data point in Fig.~2a, we numerically obtain the ground state energy density as a function of $h_2$ for fixed $h_1$ and computed its second order derivative. The phase boundary points are identified from sharp peaks.

The simulation of our QCNN in Fig.~2b also utilizes matrix product state representations. 
We first obtain the input ground state wavefunction using finite-size DMRG~\cite{McCulloch08} with bond dimension $D=130$ for a system of $N=135$ qubits. Then, the circuit operations are performed by sequentially applying SWAP and two-qubit gates on nearest neighboring qubits~\cite{Vidal04}. Each three-qubit gate is decomposed into two-qubit unitaries~\cite{Nielsen00}.
We find that increasing bond dimension to $D=150$ does not lead to any visible changes in our main figures, confirming a reasonable convergence of our method.
The color plot in Fig. 2a is similarly generated for a system of $N=45$ qubits.

\section*{QCNN for the $S=1$ Haldane Chain}

As discussed in the main text, the (spin-$1/2$) 1D cluster state belongs to an SPT phase protected by $\Z_2 \times \Z_2$ symmetry, a phase which also contains the celebrated $S=1$ Haldane chain\cite{Haldane83}. It is thus natural to ask whether this circuit can be used to detect the phase transition between the Haldane phase and an $S=1$ paramagnetic phase, which we numerically demonstrate here.

\begin{figure}[h]
           \includegraphics[width=\textwidth]{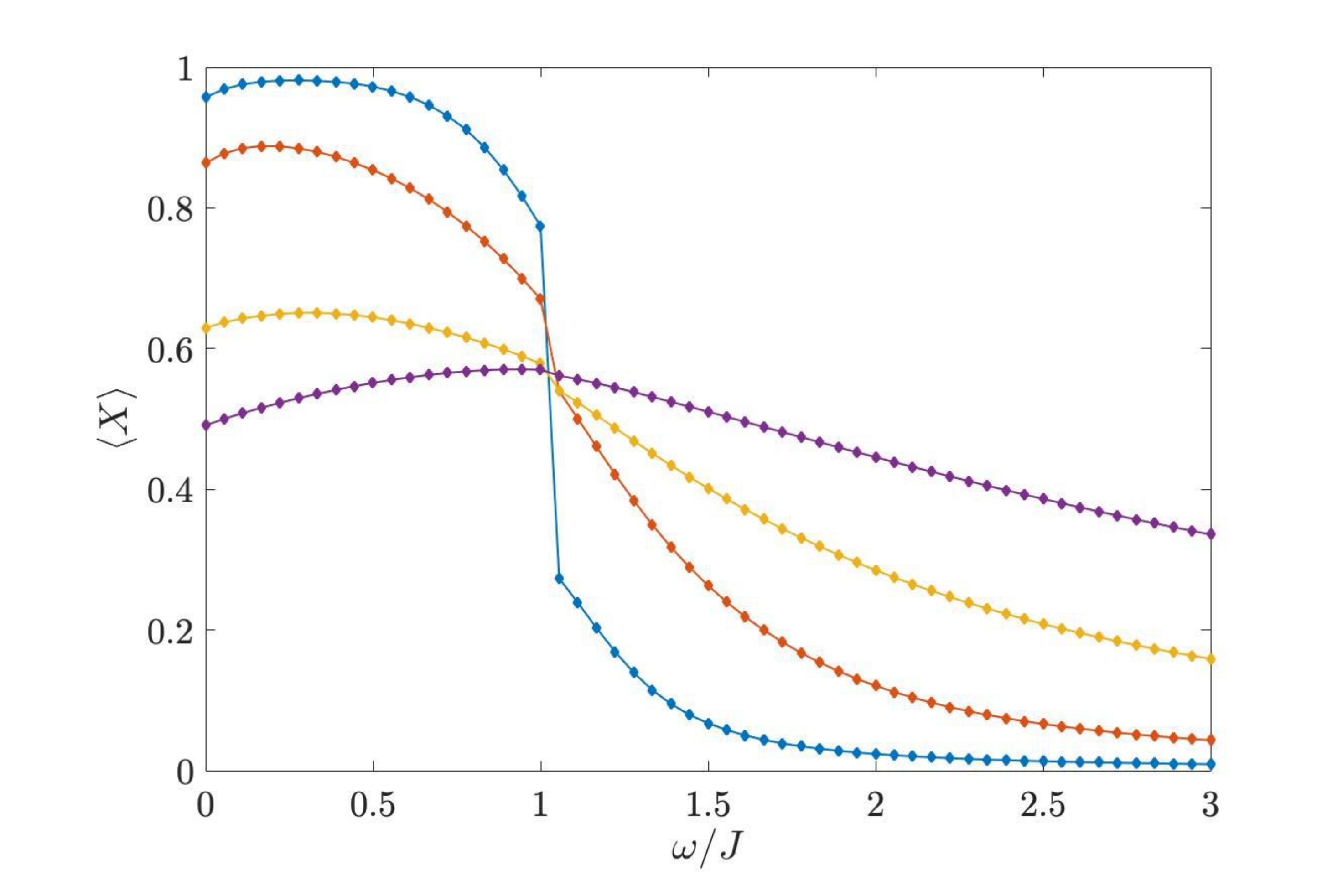}
    \caption{Exact QCNN output (at depth $d = 1, ... 4$) for the Haldane chain Hamiltonian with $N = 54$ spins.}
    \label{fig:qcnn-haldane}
\end{figure}

The one-parameter family of Hamiltonians we consider for the Haldane phase is defined on a one-dimensional chain of $N$ spin-1 particles with open boundary conditions\cite{Haldane83}: 
\begin{equation}
\label{eq:haldane}
H_{\text{Haldane}} = J \sum_{j=1}^N \mathbf{S_j} \cdot \mathbf{S_{j+1}} + \omega \sum_{j=1}^N (S_j^z)^2
\end{equation}
\noindent
In this equation, $\mathbf{S_j}$ denotes the vector of $S=1$ spin operators at site $j$. The system is protected by a $\Z_2 \times \Z_2$ symmetry generated by global $\pi$-rotations of every spin around the $X$ and $Y$ axes: $R_x = \prod_j e^{i \pi S_j^x}$, $R_y = e^{i \pi S_j^y}$. When $\omega$ is zero or small compared to $J$, the ground state belongs to the SPT phase, but when $\omega/J$ is sufficiently large, the ground state becomes paramagnetic\cite{Haldane83}. 

To apply our QCNN circuit to this Haldane phase, we must first identify a quasi-local isometric map $U$ between the two models, because their representations of the symmetry group are distinct. More specifically, since the cluster model has a $\Z_2 \times \Z_2$ symmetry generated by $X_\textrm{even(odd)} = \prod_{i \in \textrm{even(odd)}} X_i$, we require $U R_x U^\dagger = X_\textrm{odd}$ and $U R_y U^\dagger = X_\textrm{even}$. Such a map can be constructed following Ref. \onlinecite{Verresen17}. Intuitively, it extends the local Hilbert space of a spin-$1$ particle by introducing a spin singlet state $\ket{s}$ and mapping it to a pair of spin-$1/2$ particles: $\ket{x} \mapsto \ket{+-}$, $\ket{y} \mapsto -\ket{-+}$, $ \ket{z} \mapsto -i\ket{--}$, $\ket{s} \mapsto \ket{++}$.
Here, $\ket{\pm}$ denote the $\pm 1$ eigenstates of the (spin-$1/2$) Pauli matrix $X$. $\ket{\mu}$ denotes a spin-$1$ state defined by $R_\nu \ket{\mu} = (-1)^{\delta_{\mu,\nu}+1} \ket{\mu}$ ($\mu,\nu \in \{x,y,z\}$).
The QCNN circuit for the Haldane chain thus consists of applying $U$ followed by the circuit presented in the main text. 

Figure \ref{fig:qcnn-haldane} shows the QCNN output for an input system of $N = 54$ spin-1 particles at depths $d = 1,...,4$, obtained using matrix product state simulations with bond dimension $D = 160$. For this system size, we numerically identified the critical point as $\omega/J = 1.035 \pm 0.005$, by using DMRG to obtain the second derivative of energy density as a function of $\omega$ and $J$. The QCNN provides accurate identification of the phase transition.

\begin{figure}
           \includegraphics[width=0.9\textwidth]{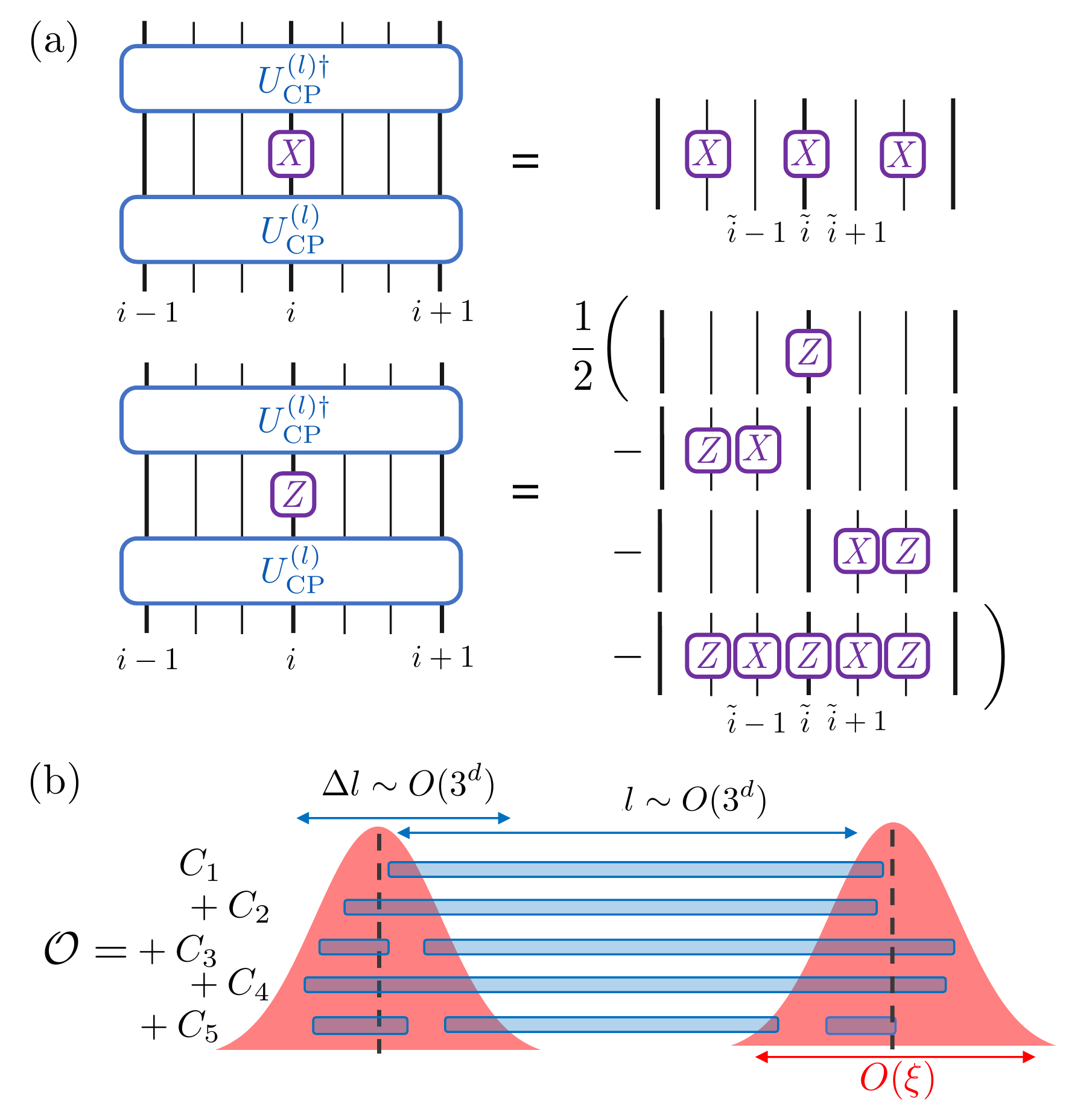}
    \caption{(a) Recursion relations for computing Pauli operators in the Heisenberg picture. $U_\text{CP}^{(l)}$ is the unitary corresponding to the convolution-pooling unit at depth $l$, and the different indices $i$, $\tilde{i}$ reflect different numbers of unmeasured qubits at different layers. Qubits measured at depth $l$ are indicated by lighter lines, while the remaining ones are shown in bold. (b) Measured operator in the Heisenberg picture is a sum of exponentially many products of string operators, with coefficients determined by Eq. (\ref{eq:multiscale-sop-sum}).}
    \label{fig:multiscale-sop}
\end{figure}

\subsection*{Multiscale String Order Parameters}
We examine the final operator measured by our circui that recognizes the SPT phase in the Heisenberg picture. Although a QCNN performs non-unitary measurements in the pooling layers, similar to QEC circuits\cite{Preskill98}, one can postpone all measurements to the end and replace pooling layers by unitary controlled gates acting on both measured and unmeasured qubits. In this way, the QCNN is equivalent to measuring a non-local observable
\begin{equation}
\label{eq:heisenberg-o}
\mathcal{O} 
=  (U_{\text{CP}}^{(d)}...U_{\text{CP}}^{(1)})^\dagger Z_{i-1} X_{i} Z_{i+1} (U_{\text{CP}}^{(d)}...U_{\text{CP}}^{(1)})
\end{equation}
where $i$ is the index of the measured qubit in the final layer and $U_{\text{CP}}^{(l)}$ is the unitary corresponding to the convolution-pooling unit at depth $l$.
A more explicit expression of $\mathcal{O}$ can be obtained by commuting $U_{\text{CP}}$ with the Pauli operators, which yields recursive relations:
\begin{equation}
\label{eq:heisenberg-x}
U_{\text{CP}}^\dagger X_{i} U_{\text{CP}} = 
X_{\tilde{i}-2} X_{\tilde{i}} X_{\tilde{i}+2}
\end{equation}
\begin{equation}
\label{eq:heisenberg-z}
\begin{split}
U_{\text{CP}}^\dagger Z_{i} U_{\text{CP}} =
\frac{1}{2} (&Z_{\tilde{i}} - Z_{\tilde{i}-2} X_{\tilde{i}-1}- X_{\tilde{i}+1} Z_{\tilde{i}+2}\\
& - Z_{\tilde{i}-2} X_{\tilde{i}-1} Z_{\tilde{i}} X_{\tilde{i}+1} Z_{\tilde{i}+2})
\end{split}
\end{equation}
$\tilde{i}$ enumerates every qubit at depth $l-1$, including those measured in the pooling layer (Fig. \ref{fig:multiscale-sop}a). It follows that an SOP of the form $ZXX...XZ$ at depth $l$ transforms into a weighted linear combination of 16 products of SOPs at depth $l-1$.
Thus, instead of measuring a single SOP, our QCNN circuit measures a sum of products of exponentially many different SOPs (Fig. \ref{fig:multiscale-sop}b): 
\begin{equation}
\label{eq:multiscale-sop-sum}
\mathcal{O} = \sum_{ab} C_{ab}^{(1)} \mathcal{S}_{ab} + \sum_{a_1 b_1 a_2 b_2} C_{a_1b_1 a_2 b_2}^{(2)} \mathcal{S}_{a_1b_1} \mathcal{S}_{a_2b_2} + \cdots,
\end{equation}
$\mathcal{O}$ can be viewed as a {\it multiscale} string order parameter with coefficients computed recursively in $d$ using Eqs. (\ref{eq:heisenberg-x},\ref{eq:heisenberg-z}). This allows the QCNN to produce a sharp classification output even when the correlation length is as long as $3^d$.

\begin{figure}
\includegraphics[width=0.45\textwidth]{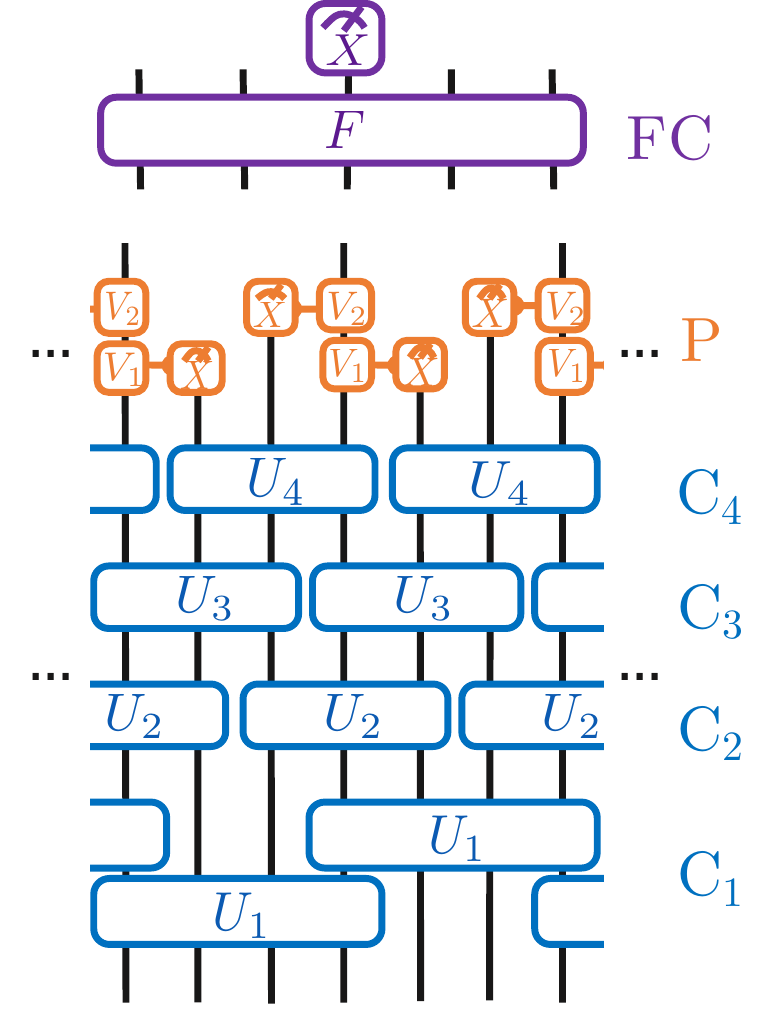}
\caption{Circuit parameterization for training a QCNN to solve QPR. Our circuit involves 4 different convolution layers ($C_1-C_4$), a pooling layer, and a fully connected layer. The unitaries are initialized to random values, and learned via gradient descent.}
\label{fig:circuit-param}
\end{figure}

\section*{Demonstration of Learning Procedure for QPR}

To perform our learning procedure in a QPR problem, we choose the hyperparameters for the QCNN as shown in Fig. \ref{fig:circuit-param}. This hyperparameter structure can be used for generic (1D) phases, and is characterized by a single integer $n$ that determines the reduction of system size in each convolution-pooling layer, $L\rightarrow L/n$. (Fig.~\ref{fig:circuit-param} shows the special case where $n=3$). The first convolution layer involves $(n+1)$-qubit unitaries starting on every $n^{\text{th}}$ qubit.
This is followed by $n$ layers of $n$-qubit unitaries arranged as shown in Fig.~\ref{fig:circuit-param}.
The pooling layer measures $n-1$ out of every contiguous block of $n$ qubits; each of these is associated with a unitary $V_j$ applied to the remaining qubit, depending on the measurement outcome.
This set of convolution and pooling layers is repeated $d$ times, where $d$ is the QCNN depth. Finally, the fully connected layer consists of an arbitrary unitary on the remaining $N/n^d$ qubits, and the classification output is given by the measurement output of the middle qubit (or any fixed choice of one of them).
For our example, we choose $n = 3$ because the Hamiltonian in Eq.~(2) involves three-qubit terms. 

In our simulations, we consider only $N = 15$ spins and depth $d = 1$, because simulating quantum circuits on classical computers requires a large amount of resources. We parameterize unitaries as exponentials of generalized $a \times a$ Gell-Mann matrices $\{\Lambda_i \}$, where $a = 2^w$ and $w$ is the number of qubits involved in the unitary\cite{Bertlmann08}: $U = \exp{\left( -i \sum_j c_j \Lambda_j \right)}$.

This parameterization is used directly for the unitaries in the convolution layers $C_2 - C_4$, the pooling layer, and the fully connected layer.
For the first convolution layer $C_1$, we restrict the choice of $U_1$ to a product of six two-qubit unitaries between each possible pair of qubits: $U_1 = U_{(23)} U_{(24)} U_{(13)} U_{(14)} U_{(12)} U_{(34)}$, where $U_{(\alpha\beta)}$ is a two-qubit unitary acting on qubits indexed by $\alpha$ and $\beta$.
Such a decomposition is useful when considering experimental implementation. 

In the QCNN learning procedure, all parameters $c_\mu$ are set to random values between $0$ and $2\pi$ for the unitaries $\{ U_i, V_j, F \}$. In every iteration of gradient descent, we compute the derivative of the mean-squared error function (Eq. (1) in the main text) to first order with respect to each of these coefficients $c_\mu$ by using the finite-difference method:
\begin{equation}
\frac{\partial \text{MSE}}{\partial c_\mu} = \frac{1}{2 \epsilon} \left(\text{MSE}(c_\mu+\epsilon) - \text{MSE}(c_\mu-\epsilon) \right) + O(\epsilon^2).
\end{equation}
Each coefficient is thus updated as $c_\mu \mapsto c_\mu - \eta \frac{\partial \text{MSE}}{\partial c_\mu}$, where $\eta$ is the learning rate for that iteration. We compute the learning rate using the bold driver technique from machine learning, where $\eta$ is increased by 5\% if the error has decreased from the previous iteration, and decreased by 50\% otherwise~\cite{Hinton07}. We repeat the gradient descent procedure until the error function changes on the order of $10^{-5}$ between successive iterations. In our simulations, we use $\epsilon = 10^{-4}$ for the gradient computation, and begin with an initial learning rate of $\eta_0 = 10$.

\section*{Construction of QCNN Circuit}

To construct the exact QCNN circuit in Fig. 2b, we followed the guidelines discussed in the main text. Specifically, we designed the convolution and pooling layers to satisfy the following two important properties:

\begin{enumerate}
\item
Fixed-point criterion: If the input is a cluster state $\ket{\psi_0}$ of $L$ spins, the output of the convolution-pooling layers is a cluster state $\ket{\psi_0}$ of $L/3$ spins, with all measurements deterministically yielding $\ket{0}$.
\item 
QEC criterion: If the input is not $\ket{\psi_0}$ but instead differs from $\ket{\psi_0}$ at one site by an error which commutes with the global symmetry, the output should still be a cluster state of $L/3$ spins, but at least one of the measurements will result in the state $\ket{1}$.
\end{enumerate}

\noindent
These two properties are desirable for any quantum circuit implementation of RG flow for performing QPR.

In the specific case of our Hamiltonian, the ground state (1D cluster state) is a graph state, which can be efficiently obtained by applying a sequence of controlled phase gates to a product state.
This significantly simplifies the construction of the MERA representation for the fixed-point criterion.
To satisfy the QEC criterion, we treat the ground state manifold of the unperturbed Hamiltonian $H = -J \sum_i Z_i X_{i+1} Z_{i+2}$ as the code space of a stabilizer code with stabilizers $\{Z_i X_{i+1} Z_{i+2} \}$. The remaining degrees of freedom in the QCNN convolution and pooling layers are then specified such that the circuit detects and corrects the error (i.e. measures at least one $\ket{1}$ and prevents propagation to the next layer) when a single-qubit $X$ error is present.

\begin{figure}
\includegraphics[width=0.9\textwidth]{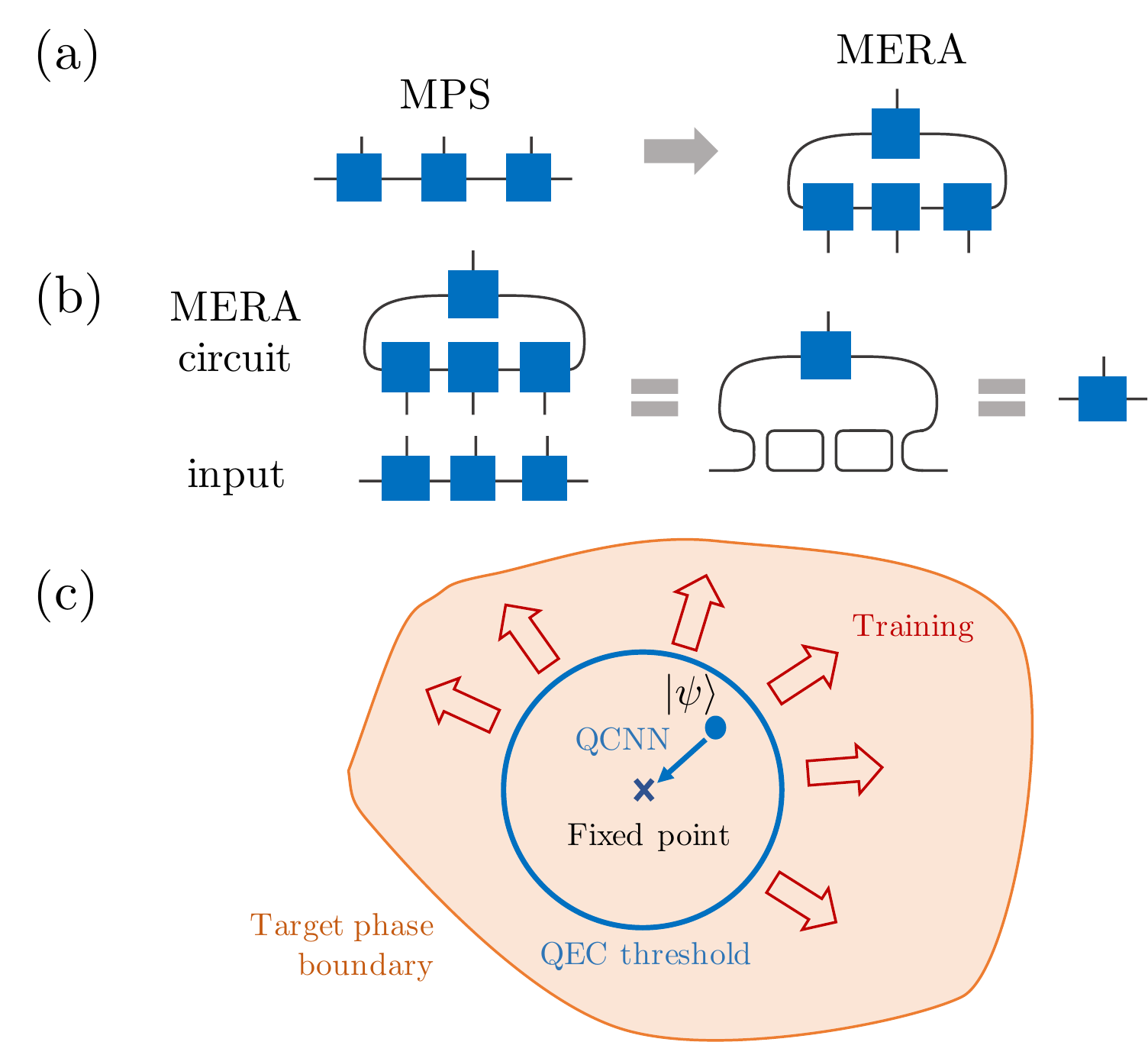}
\caption{(a) Given a state with a translationally invariant, isometric matrix product state representation (e.g. a fixed point state for a 1D SPT phase), we explicitly construct an isometry for the MERA representation of this state. Blue squares are the matrix product state tensors, while black lines are the legs of the tensor.  While we have illustrated a 3-to-1 isometry, the generalization to arbitrary $n$-to-1 isometries is straightforward. (b) Diagrammatic proof showing that a MERA constructed from the above tensor maps the fixed-point state back to a shorter version of itself. The first equality uses the definition of isometric tensor, and loops in the middle diagram simplify to a constant number unity. The generalization of this isometry to higher dimensions is discussed in Ref. \onlinecite{Schuch10}.
(c) One helpful initial parameterization for QPR problems consists of a MERA for the fixed point state $\ket{\psi_0 (\mathcal{P})}$ and a choice of nested QEC, so that states within the QEC threshold flow toward $\ket{\psi_0 (\mathcal{P})}$.
Training procedures then expand this threshold boundary to the phase boundary.}
\label{fig:generalizations}
\end{figure}

\section*{QCNN for General QPR Problems}

Our interpretation of QCNNs in terms of MERA and QEC motivates their application for recognizing more generic quantum phases.
For any quantum phase $\mathcal{P}$ whose RG fixed-point wavefunction $\ket{\psi_0 (\mathcal{P})}$ has a tensor network representation in isometric or $G$-isometric form~\cite{Schuch11} (Fig. \ref{fig:generalizations}a), one can systematically construct a corresponding QCNN circuit.
This family of quantum phases includes all 1D SPT and 2D string-net phases~\cite{Chen11,Schuch11,Levin04}. 
In these cases, one can explicitly construct a commuting parent Hamiltonian for $\ket{\psi_0 (\mathcal{P})}$ and a MERA structure in which $\ket{\psi_0(\mathcal{P})}$ is a fixed-point wavefunction (Fig.~\ref{fig:generalizations}a for 1D systems). tThe diagrammatic proof of this fixed-point property is given in Fig.~\ref{fig:generalizations}b. 
Furthermore, any ``local error'' perturbing an input state away from $\ket{\psi_0 (\mathcal{P})}$ can be identified by measuring a fraction of terms in the parent Hamiltonian, similar to syndrome measurements in stabilizer-based QEC\cite{Preskill98}.
Then, a QCNN for $\mathcal{P}$ simply consists of the MERA for $\ket{\psi_0 (\mathcal{P})}$ and a nested QEC scheme in which an input state with error density below the QEC threshold~\cite{Aharonov97} ``flows'' to the RG fixed point. Such a QCNN can be optimized via our learning procedure.

While our generic learning protocol begins with completely random unitaries, as in the classical case\cite{LeCun15}, this initialization may not be the most efficient for gradient descent. Instead, motivated by deep learning techniques such as pre-training\cite{LeCun15}, a better initial parameterization would consist of a MERA representation of $\ket{\psi_0 (\mathcal{P})}$ and one choice of nested QEC. With such an initialization,
the learning procedure serves to optimize the QEC scheme, expanding its threshold to the target phase boundary (Fig. \ref{fig:generalizations}c). 

\section*{Experimental Resource Analysis}

To compute the gate depth of the cluster model QCNN circuit in a Rydberg atom implementation, we analyze each gate shown in Figure \ref{fig:cluster-qcnn}b. By postponing pooling layer measurements to the end of the circuit, the multi-qubit gates required are
\begin{equation}
\text{C}_z\text{Z}_{ij} = e^{i \pi (-\one+Z_i) (-\one+Z_j)/4 }
\end{equation}
\begin{equation}
\text{C}_x\text{Z}_{ij} = e^{i \pi (-\one+X_i) (-\one+Z_j)/4 }
\end{equation}
\begin{equation}
\text{C}_x\text{C}_x\text{X}_{ijk} = e^{i \pi (-\one+X_i) (-\one+X_j) (-\one+X_k)/8}.
\end{equation}
By using Rydberg blockade-mediated controlled gates\cite{Saffman10}, it is straightforward to implement $\text{C}_z\text{Z}_{ij}$ and $\text{C}_z\text{C}_z\text{Z}_{ijk} = e^{i \pi (-\one+Z_i) (-\one+Z_j) (-\one+Z_k)/8}$. The desired $\text{C}_x\text{Z}_{ij}$ and $\text{C}_x\text{C}_x\text{X}_{ijk}$ gates can then be obtained by conjugating $\text{C}_z\text{Z}_{ij}$ and $\text{C}_z\text{C}_z\text{Z}_{ijk}$ by single-qubit rotations. 
For input size of $N$ spins, the $k^{\text{th}}$ convolution-pooling unit thus applies ${4N}/{3^{k-1}}$ $\text{C}_z\text{Z}_{ij}$ gates, $N/3^{k-1}$ $\text{C}_x\text{C}_x\text{X}_{ijk}$ gates, and ${2N}/{3^{k-1}}$ layers of $\text{C}_x\text{Z}_{ij}$ gates. The depth of single-qubit rotations required is $4d$, as these rotations can be implemented in parallel on all $N$ qubits.
Finally, the fully connected layer consists of $N 3^{1-d}$ $\text{C}_z\text{Z}_{ij}$ gates. Thus, the total number of multi-qubit operations required for a QCNN of depth $d$ operating on $N$ spins is $\frac{7N}{2}(1-3^{1-d}) + N 3^{1-d}$. Note that we need not use SWAP gates since the Rydberg interaction is long-range.

\section*{Demonstration of Learning Procedure for QEC}

To obtain the QEC code considered in the main text, we consider a QCNN with $N = 9$ input physical qubits and simulate the circuit evolution of its $2^N \times 2^N$ density matrix exactly.
Strictly speaking, our QCNN has three layers: a three-qubit convolution layer $U_1$, a 3-to-1 pooling layer, and a 3-to-1 fully connected layer $U_2$. 
 Without loss of generality, we may ignore the optimization over the pooling layer by absorbing its effect into the first convolution layer, leading to the effective two-layer structure shown in Fig. \ref{fig:qcnn-qec}a.
 The  generic three-qubit  unitary operations $U_1$ and $U_2$ are parameterized using 63 Gell-Mann coefficients each.

\begin{figure}
\includegraphics[width=0.65\textwidth]{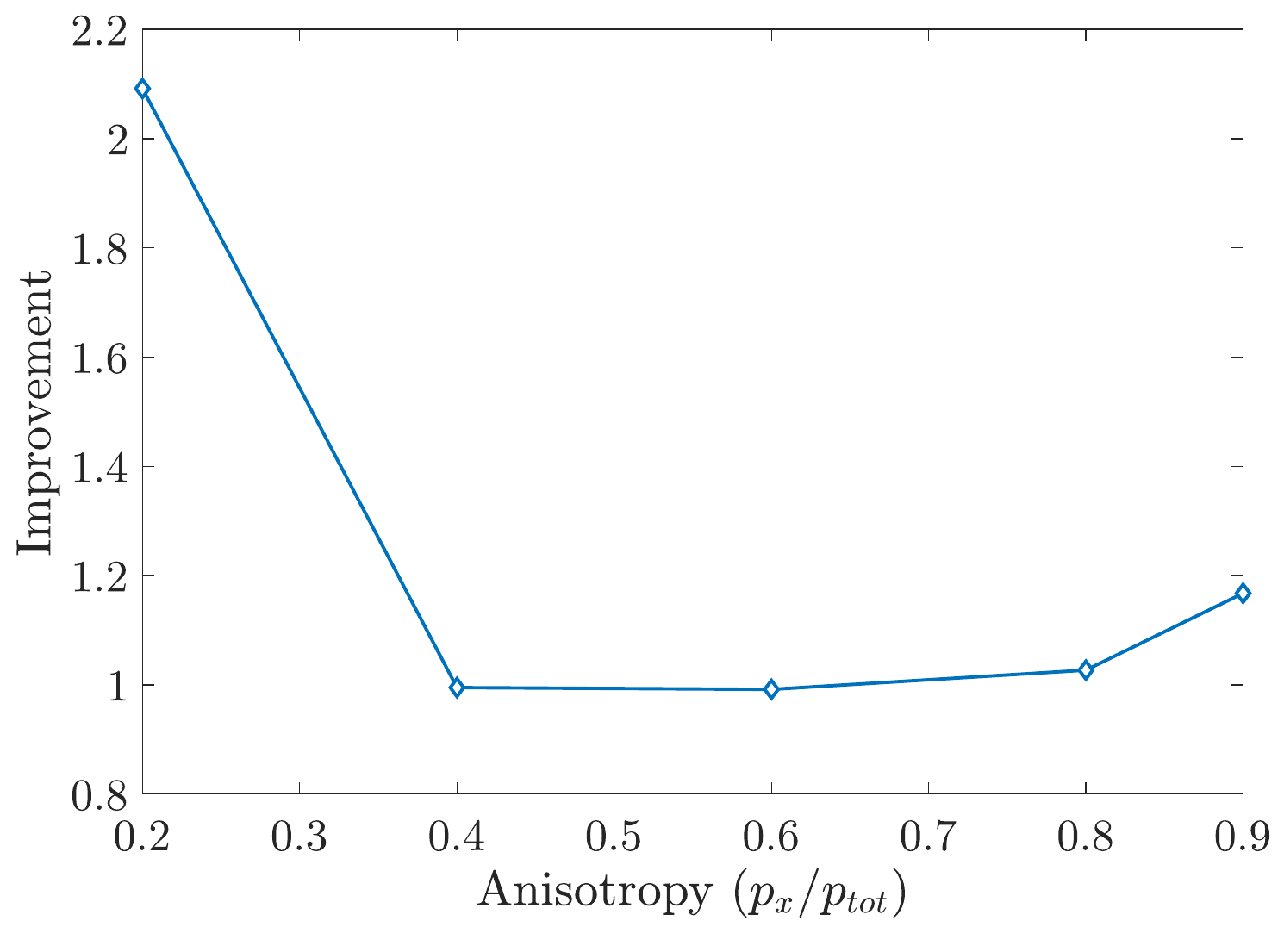}
\caption{Ratio between the logical error rate of the Shor code and that of the QCNN code for the anisotropic depolarization error model. We fix the total input error rate $p_{\text{tot}} = p_x+p_y+p_z = 0.001$ and $p_y = p_z$, while varying the ratio $p_x / p_{\text{tot}}$.}
\label{fig:qec-anisotropy}
\end{figure}

As discussed in the main text, we consider three different error models: (1) independent single-qubit errors on all qubits with equal probabilities $p_\mu$ for $\mu=X$, $Y$, and $Z$ errors, (2) 
independent single-qubit errors on all qubits, with 
anisotropic probabilities $p_x \neq p_y = p_z$, and (3) independent single-qubit anisotropic errors with additional two-qubit correlated errors $X_i X_{i+1}$ with probability $p_{xx}$. 
More specifically, the first two error models are realized by applying a (generally anisotropic) depolarization quantum channel to each of the nine physical qubits:
\begin{equation}
\mN_{1,i}: \rho \mapsto (1- \sum_\mu p_\mu) \rho +\sum_{\mu} p_\mu \sigma^\mu_i \rho \sigma^\mu_i
\end{equation}
with Pauli matrices $\sigma^\mu_i$ for $i\in \{1, 2, \dots, 9\}$ (the qubit indices are defined from bottom to top in Fig.~\ref{fig:qcnn-qec}a).
For the anisotropic case, we trained the QCNN on various different error models with the same total error probability $p_x+p_y+p_z=0.001$, but different relative ratios; the resulting ratio between the logical error probability of the Shor code and that of the QCNN code is plotted as a function of anisotropy in Fig. \ref{fig:qec-anisotropy}. For strongly anisotropic models, the QCNN outperforms the Shor code, while for nearly isotropic models, the Shor code is optimal and QCNN can achieve the same logical error rate.

For the correlated error model, we additionally apply a quantum channel:
\begin{equation}
\mN_{2,i}: \rho \mapsto (1-p_{xx}) \rho + p_{xx} X_i X_{i+1} \rho X_i X_{i+1}
\end{equation}
for pairs of nearby qubits, i.e. $i \in \{1,2,4,5,7,8\}$.
Such a geometrically local correlation is motivated from experimental considerations. In this case,
we train our QCNN circuit on a specific error model with parameter choices $p_x = 5.8 \times 10^{-3}$, $p_y = p_z = 2 \times 10^{-3}$, $p_{xx} = 2 \times 10^{-4}$ and evaluate the logical error probabilities for various physical error models with the same relative ratios, but different total error per qubit $p_x+p_y+p_z+p_{xx}$. In general, for an anisotropic logical error model with probabilities $p_\mu$ for $\sigma_\mu$ logical errors, the overlap $f_q$ is $(1-2\sum_\mu p_\mu/3)$, since $\bra{\pm \nu} \sigma_\mu \ket{\pm \nu} = (-1)^{\delta_{\mu,\nu}+1}$. Becuase of this, we compute the total logical error probability from $f_q$ as $1.5(1-f_q)$. 
Hence, our goal is to maximize the logical state overlap $f_q$ defined in Eq. (\ref{eq:qcnn-overlap}).
If we naively apply the gradient descent method based on $f_{q}$ directly to both $U_1$ and $U_2$, we find that the optimization is easily trapped in a local optimum.
Instead, we optimize two unitaries $U_1$ and $U_2$ sequentially, similar to the layer-by-layer optimization in backpropagation for conventional CNN\cite{LeCun15}.

A few remarks are in order. First, since $U_1$ is optimized prior to $U_2$, one needs to devise an efficient cost function $C_1$ that is independent of $U_2$. In particular, simply maximizing $f_{q}$ with an assumption $U_2 = \mathbf{1}$ may not be ideal, since such choice does not capture a potential interplay between $U_1$ and $U_2$.
Second, because $U_1$ captures arbitrary single qubit rotations, the definition of $C_1$ should be basis independent. 
Finally, we note that the tree structure of our circuit allows one to view the first layer as an independent quantum channel:
\begin{equation}
\mM_{U_1}: \rho \mapsto \textrm{tr}_{a}[ U_1 \mN (U_1^\dagger (\ket{0}\bra{0} \otimes \rho \otimes \ket{0}\bra{0}) U_1) U_1^\dagger],
\end{equation}
where $\textrm{tr}_{a}[ \cdot ]$ denotes tracing over the ancilla qubits that are measured in the intermediate step.
From this perspective, $\mM_{U_1}$ describes an effective error model to be corrected by the second layer.

With these considerations, we optimize $U_1$ such that the effective error model $\mM_{U_1}$ becomes as classical as possible, i.e. $\mM_{U_1}$ is dominated by a ``flip'' error along a certain axis with a strongly suppressed ``phase'' error.
Only then, the remant, simpler errors will be corrected by the second layer.
More specifically, one may represent $\mM_{U_1}$ using a map $M_{U_1}: \mathbf{r}\mapsto M \mathbf{r} + \mathbf{c}$, where $\mathbf{r}\in \mathbb{R}^3$ is the Bloch vector for a qubit state $\rho \equiv\frac{1}{2} \mathbf{1}+ \mathbf{r} \cdot \mathbf{\sigma}$~\cite{Nielsen00}. The singular values of the real matrix $M$ encode the probabilities $p_1\geq p_2\geq p_3$ for three different types of errors.
We choose our cost function for the first layer as $C_1 = p_1^2 + p_2 + p_3$, which is relatively more sensitive to $p_2$ and $p_3$ than $p_1$ and ensure that the resultant, optimized channel $\mM_{U_1}$ is dominated by one type of error (with probability $p_1$).
We note that $M$ can be efficiently evaluated from a quantum device without knowing $\mathcal{N}$, by performing quantum process tomography for a single logical qubit.
Once $U_1$ is optimized, we use gradient decent to find an optimal $U_2$ to maximize the fidelity $f_q$.  
As with QPR, gradients are computed via the finite-difference method, and the learning rate is determined by the bold driver technique\cite{LeCun15}.

\end{document}